\documentclass[aps,prl,twocolumn,floatfix]{revtex4}
\usepackage{amssymb}
\usepackage{amsmath}
\usepackage{graphicx}

\begin{document}

\title{Phase of quantum oscillation in Weyl semimetals}

\author{G.~P.~Mikitik}
\affiliation{B.~Verkin Institute for Low Temperature Physics \&
Engineering, Ukrainian Academy of Sciences, Kharkiv 61103,
Ukraine}

\author{Yu.~V.~Sharlai}
\affiliation{B.~Verkin Institute for Low Temperature Physics \&
Engineering, Ukrainian Academy of Sciences, Kharkiv 61103,
Ukraine}

\begin{abstract}
We consider the semiclassical quantization condition for the energy of an electron in a magnetic field in the case when the electron orbit lies on a Fermi-surface pocket surrounding the Weyl point of a topological semimetal and analyze the constant $\gamma$ appearing in this condition. It is shown that this constant has the universal value, $\gamma=0$, independent of the tilt of the Weyl spectrum. Since the constant $\gamma$ for an extremal cross section of the Fermi surface determines the phase of quantum oscillations, this result explains why measurements of the phase permit one to find  Weyl points in crystals even though the extremal cross section of the pocket does not pass through this point, and the appropriate Berry phase of the orbit differs from $\pi$.
\end{abstract}

\maketitle

\begin{figure}[bp] 
 \centering  \vspace{+9 pt}
\includegraphics[scale=1.]{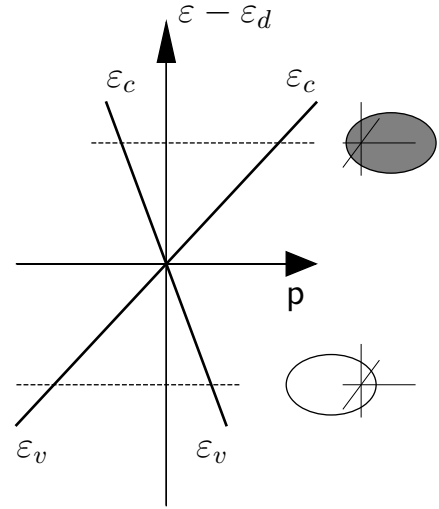}
\caption{\label{fig1} Dispersion relations $\varepsilon_c(p)$ and $\varepsilon_v(p)$ of the conduction and valent energy bands in the vicinity of the degeneracy (Weyl) point in the case of $\tilde a^2<1$. On the right the Fermi-surface ellipsoids at $\zeta-\varepsilon_d<0$ and $\zeta-\varepsilon_d>0$ are shown together with the degeneracy point which is the origin of the coordinate axes. Note the shift of the centers of the ellipsoids relative to the origin. The shaded and white surfaces correspond to the electron and hole charge carriers, respectively.
 } \end{figure}

The topological Weyl semimetals have attracted much attention in recent years \cite{armit}. In these semimetals, two electron energy bands, which are not degenerate in electron spin, contact at discrete Weyl points of the Brillouin zone and disperse linearly in all directions around these specific points (Fig.~\ref{fig1}). This type of the band degeneracy can occur in crystals with strong spin-orbit interaction when either the spatial inversion or time reversal symmetry is broken. Below we shall consider only the crystals without the inversion symmetry. This class of the Weyl semimetals includes, e.g.,  TaAs, TaP, NbAs, NbP, in which the Fermi energy $\varepsilon_F$ lies near the Weyl points of these crystals.

Any extremal electron orbit on the Fermi surface of a metal produces an oscillation contribution to the magnetization or conductivity \cite{LP}, and this contribution is a periodic function of the following phase \cite{LP,shen}:
 \[
 \frac{F}{H}-\gamma,
 \]
where $F=cS_{ex}/(2\pi e\hbar)$ is the quantum-oscillation frequency corresponding to the extremal cross-sectional area $S_{ex}$, $\gamma$ is the constant in the well-known semiclassical quantization condition for the electron energy $\varepsilon_l(p_z)$ in the magnetic field $H$ \cite{LP,Sh},
\begin{equation}\label{1}
S(\varepsilon_{l},p_z)=\frac{2\pi\hbar e H}{c}\left(l+\gamma\right),
\end{equation}
$l=0,1,...$ is an integer, $p_z$ is the quasimomentum in the direction of the magnetic field, and $S(\varepsilon,p_z)$ is the area of a Fermi-surface cross section perpendicular to the magnetic field. Thus, a value of the constant $\gamma$ (or electron $g$ factor in crystals with inversion symmetry) determines the offset of the quantum-oscillation phase, and measurements of this offset in the Shubnikov - de Haas and de Haas - van Alphen effects permit one to find those electron states that lead to a special value of $\gamma$. In particular, one of the methods of detecting degenerate electron states in topological semimetals is just based on such measurements (see, e.g., numerous references in review \cite{m-sh19}).

The value of $\gamma$ was investigated in many papers, see, e.g.,  Refs.~\cite{zilb,azb,falk,slut,jetp,prl,g1,g2,alex}. Here we mention the two situations when the constant $\gamma$ changes as compared to the case of the conventional electrons, the dispersion of which contains neither a degeneracy point nor a band-contact line. i) At weak spin-orbit interaction when the degeneracy of the bands occurs along lines in the Brillouin zone (this case occurs, e.g., in nodal-line semimetals \cite{armit}), the value of $\gamma$ for the orbits that are penetrated by the line  ($\gamma=0$) \cite{jetp,prl} differs from the value $\gamma=1/2$ \cite{zilb,LP} characteristic of the conventional quasiparticles. This difference in $\gamma$ results from the nonzero Berry phase for the orbits surrounding the band-contact lines. ii) At strong spin-orbit interaction in crystals with the inversion symmetry, the degeneracy of the bands can occurs at certain  points of the Brillouin zone (Dirac semimetals just fall into this class \cite{armit}). It turned out that in this case, for the orbits in the vicinity of such Dirac points, the electron $g$ factor takes on  a specific value, and the appropriate quantization condition, in fact, reduces to Eq.~(\ref{1}) with $\gamma=0$ \cite{g1,g2}. This specific value of $\gamma$ is now associated both with the nonzero Berry phase of the electron orbit and with a certain part of the electron orbital moment. The above-mentioned results demonstrate that  measurements of the phase offset  do make it possible to detect the degeneracy lines and points in the nodal-line and Dirac semimetals. In this paper, we shall analyze the value of $\gamma$ for the Weyl semimetals.

At a strong spin-orbit interaction in a crystal without a center of inversion, the constant $\gamma$ for any electron orbit in the Brillouin zone  is determined by the formula  \cite{fuchs,m-sh12,alex}:
\begin{equation} \label{2}
\gamma -\frac{1}{2}=-\frac{\Phi_B}{2 \pi} -\frac{1}{2 \pi\hbar m } \oint_{\Gamma} \frac{L({\bf p})} {v_{\perp}({\bf p})} d{\bf \kappa},
\end{equation}
where $\Phi_B$ is the Berry phase of this orbit $\Gamma$, $v_{\perp}$ is the absolute value of projection of the electron velocity ${\bf v}= \partial \varepsilon({\bf p})/\partial {\bf
p}$  on the plane of the orbit, $\varepsilon({\bf p})$ is the dispersion of the charge carriers, $m$ is electron mass, $d\kappa$ is the infinitesimal element of the orbit, and $L$ is those part of the electron orbital moment that is normal to the plane of the orbit and that is  associated with self-rotation of the semiclassical wave packet around its center of mass \cite{niu}. In general case, this   $\gamma$ can take on any  value.

Consider now the case of the Weyl point. The most general ${\bf k}\cdot{\bf p}$ Hamiltonian $\hat H$ for the conduction and valence electron bands in the vicinity of the Weyl point has the form \cite{m-sh19}:
\begin{eqnarray} \label{3}
\hat H=(\varepsilon_d+h_0({\bf p}))\hat\sigma_0\!+\!\sum_{i=1}^3\!h_i({\bf
p})\hat\sigma_i ,
\end{eqnarray}
where $\hat\sigma_i$ are the Pauli matrices, $\hat\sigma_0$ is the unit matrix, $h_0({\bf p})$, $h_i({\bf p})$ are linear functions of the quasimomentum: $h_0={\bf a}{\bf p}$, $h_{1}={\bf v}^{(1)}{\bf p}$, $h_{2}={\bf v}^{(2)}{\bf p}$, $h_3={\bf a}'{\bf p}$. Here ${\bf a}$, ${\bf a}'$, ${\bf v}^{(1)}$, ${\bf v}^{(2)}$ are real constant vectors,  $\varepsilon_d$ is the energy of the Weyl point, and the quasimomentum  ${\bf p}$ is measured from this point. The diagonalization of the Hamiltonian gives the dispersion of the bands $c$ and $v$,
 \begin{eqnarray}\label{4}
 \varepsilon_{c,v}({\bf p})=\varepsilon_d+{\bf a}\cdot{\bf p}\pm E({\bf p}),
 \end{eqnarray}
where  $[E({\bf p})]^2=\sum_{i=1}^3h_i^2$ is a positively definite quadratic form in the components of the vector ${\bf p}$. Let us choose the coordinate axes along principal directions of this form. In this case, one has
 \begin{eqnarray}\label{5}
[E({\bf p})]^2=b_{11}p_1^2+b_{22}p_2^2+b_{33}p_3^2,
 \end{eqnarray}
where  $b_{11}$, $b_{22}$, $b_{33}$ are the positive constants. The vector ${\bf \tilde a}$ with the components $\tilde a_i\equiv a_i/\sqrt{b_{ii}}$ characterizes the so-called tilt of the spectrum, and if
\begin{eqnarray*}
\tilde a^2 \equiv \frac{a_1^2}{b_{11}}+ \frac{a_2^2}{b_{22}} +\frac{a_3^2}{b_{33}}<1,
 \end{eqnarray*}
the Fermi surface is an ellipsoid, with its center being shifted relative to the point ${\bf p}=0$ by the vector ${\bf s}$ that is  proportional to $\varepsilon_F-\varepsilon_d$ (Fig.~\ref{fig1}). In particular, in the case when only $a_1$ and $a_2$ differ from zero, this vector has the form,
 \begin{eqnarray}\label{6}
 {\bf s}=-\frac{(\varepsilon_F-\varepsilon_d)}{1-\tilde a_1^2 -\tilde a_2^2}\left(\frac{\tilde a_1}{\sqrt{b_{11}}},\frac{\tilde a_2}{\sqrt{b_{22}}},0\right).
 \end{eqnarray}
We emphasize that the tilt of the spectrum is the inherent property of the Weyl points in a crystal since they are not highly-symmetric points of the Brillouin zone \cite{armit}. Therefore, as is clear from Fig.~\ref{fig1}, if the magnetic field ${\bf H}$ is not perpendicular to ${\bf s}$,  the appropriate maximal cross sections of the ellipsoid  do not pass through the Weyl point. At $\tilde a^2>1$, a closed Fermi surface does not exist near the Weyl point, and this case corresponds to the so-called Type-II Weyl semimetals \cite{sol}.

Considering a charge-carrier dispersion equivalent to Eq.~(\ref{4}) with ${\bf a}=0$ (Appendix C in Ref.~\cite{alex}), and using formula (\ref{2}), it was obtained  \cite{fuchs,alex} that $\gamma=0$ in this case of absence of the tilt. Below we consider the realistic  situation of ${\bf a}\neq 0$. Let us turn the coordinate system so that its $p_z$ axis coincides with the direction of the magnetic  field ${\bf n}\equiv {\bf H}/H$. Then, an electron orbit in this  magnetic field is determine by the conditions,
 \begin{eqnarray}\label{7}
 \varepsilon(p_x,p_y,p_z)=\varepsilon_F,\ \ \
 p_z={\rm const.},
 \end{eqnarray}
i.e., the plane of the orbit is parallel to the coordinate $p_x$-$p_y$ plane. In the new coordinate system, $h_0({\bf p})$, $h_i({\bf p})$ are still linear functions of the quasimomentum: $h_0={\bf a}_0{\bf p}$, $h_{i}={\bf a}_i{\bf p}$ where the constant vectors  ${\bf a}_0$, ${\bf a}_i$ are expressed in terms of ${\bf a}$, ${\bf v}^{(1)}$, ${\bf v}^{(2)}$, ${\bf a}'$, respectively. In these coordinates, the explicit expressions for  $\Phi_B$ and $L({\bf p})$ were found in Ref.~\cite{m-sh12}, using Hamiltonian (\ref{3}). For example, the expressions for the band $c$ look like:
\begin{eqnarray}\label{8}
\Phi_B&=&\!\!\frac{1}{\hbar}\oint_{\Gamma}{\bf \Omega}_c\,d{\bf p}= \frac{1}{\hbar}\oint_{\Gamma}\frac{d\kappa ({\bf \Omega}_c[{\bf n}\times {\bf v}({\bf p})])}{v_{\perp}({\bf p})}, \nonumber \\
{\bf \Omega}_c&=&\!\!\frac{\hbar}{2(h_1^2+h_2^2)}\!\left(
h_2\frac{\partial h_1}{\partial {\bf p}}-h_1\frac{\partial
h_2}{\partial {\bf p}}\right)\!\!\!\left(1-\frac{h_3}{|E({\bf p})|} \right)\!,~~ \\
  L_c &=&\!\!-\frac{\hbar \,m}{2(E({\bf p}))^2}
\sum_{i,j,k=1}^3\varepsilon_{ijk}h_i\frac{\partial h_j}{\partial
p_x} \frac{\partial h_k}{\partial p_y}, \nonumber
\end{eqnarray}
where $\varepsilon_{ijk}$ is the completely antisymmetric unit tensor with $\varepsilon_{123}=1$, and we have neglected the contribution of the relatively small Zeeman term associated with the electron spin to $L_c$. Taking into account formulas (\ref{7}), (\ref{8}), and the definitions of $h_i({\bf p})$,
the direct calculation gives,
\begin{eqnarray}\label{9}
\frac{L_c}{m\hbar}\!&=&\!-\frac{p_z({\bf a}_1[{\bf a}_2\times {\bf a}_3])}{2(\varepsilon_F\!-\!\varepsilon_d-h_0)^2},  \\
\!\!\!\!\!\!\frac{({\bf \Omega}_c[{\bf n}\!\!\times\!{\bf v}({\bf p})])}{\hbar}\!\!\!&
=&\!\!\!\frac{1}{2(\varepsilon_F\!-\!\varepsilon_d\!-\!h_0)(\varepsilon_F\! -\!\varepsilon_d\!-\!h_0\!+\!h_3)}\times  \\
\Big[\!(\varepsilon_F\!-\!\varepsilon_d)[{\bf a}_2\!\!\!\!&\times&\!\!\!{\bf a}_1]_z\!\!
+\!p_z({\bf a}_0[{\bf a}_1\!\!\!\times\!{\bf a}_2])\!+\!\frac{p_zh_3({\bf a}_1[{\bf a}_2\!\!\times\!{\bf a}_3])}{(\varepsilon_F\!-\!\varepsilon_d-h_0)} \Big]\!. \nonumber \label{10}
\end{eqnarray}

According to Eqs.~(\ref{7}), $p_y$ in the orbit is a function of $p_x$, and the integration over the orbit can be reduced to the   integration over $p_x$. Considering $p_x$ as a variable in the complex plane, all the above integrals are calculated analytically, using  the residue theorem. For the second and first terms in the right hand side of formula (\ref{2}), we obtain
\begin{eqnarray}\label{11}
\frac{p_z({\bf a}_1[{\bf a}_2\!\times\!{\bf a}_3])}{2 \sqrt{\sum_{i<j}((\varepsilon_F\!-\!\varepsilon_d)[{\bf a}_i\!\times\! {\bf a}_j]_z\!-\!p_z({\bf a}_0[{\bf a}_i\!\times\!{\bf a}_j]))^2}}\!\equiv\!\delta \gamma_L,~ \\
-\frac{\Phi_B}{2\pi}=-\frac{1}{2}-\delta\gamma_L,~ \label{12}
\end{eqnarray}
respectively. Inserting Eqs.~(\ref{11}) and (\ref{12}) into formula (\ref{2}), we arrive at the main result of our paper,
\begin{eqnarray}\label{13}
 \gamma=0.
\end{eqnarray}
Note that formula (\ref{13}) is valid at any values of the parameters if $\tilde a^2<1$, i.e., if the Fermi surface is the ellipsoid. (As was mentioned above, at $\tilde a^2>1$, a closed Fermi surface does not exist near the Weyl point.) Equality (\ref{13}) generalizes the result obtained in Refs.~\cite{fuchs,alex} to the case ${\bf a}\neq 0$ that occurs in real Weyl semimetals. The zero value of $\gamma$ means that the quantum oscillation produced by the electron orbits located near a Weyl point has the specific phase offset that differs from the offset of the conventional electrons and that permits one to detect these points in crystals. The result $\gamma=0$ also completely agrees with the exact spectrum obtained in Ref.~\cite{m-sh} for electrons  in the vicinity of a band-contact point in the magnetic field.

\begin{figure}[t] 
 \centering  \vspace{+9 pt}
\includegraphics[scale=1.]{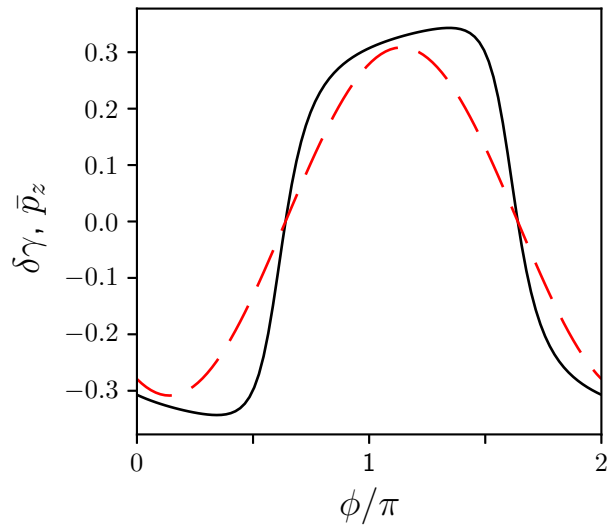}
\caption{\label{fig2} The quantity $\delta\gamma_L$, which coincides with $(\Phi_B/2\pi)-0.5$, (solid line), and the dimensionless $\bar{p}_z=p_z/p_{\rm max}$ (dashed line) versus the angle $\phi$ defining the direction of the magnetic field in the plane $p_3=0$. Here $p_{\rm max}\approx 4.8|\zeta-\varepsilon_d|/(b_{11}b_{22})^{1/4}$ is the longest axis of the elliptic cross section of the Fermi-surface pocket by this plane; $\delta\gamma_L$ is described by formula (\ref{16}) with parameters presented in Eqs.~(\ref{15}).
 } \end{figure}   

To illustrate the obtained results, let us consider the small Fermi-surface pocket located at the Weyl point W1 of TaAs \cite{arnold16}. Near this point, dispersions of the two bands  can be described by Hamiltonian (\ref{3}) with
 \begin{eqnarray}\label{14}
 {\bf a}=(a_1,a_2,0),\ \ \ {\bf a}'=(\sqrt{b_{11}},0,0),\\
 {\bf v}_1=(0,\sqrt{b_{22}},0),\ \ \ {\bf v}_2=(0,0,\sqrt{b_{33}}). \nonumber
 \end{eqnarray}
Using the experimental data of Arnold et al. \cite{arnold16}, values of all the parameters $b_{ii}$ and $a_i$ were found in Ref.~\cite{m-sh21}:
 \begin{eqnarray}\label{15}
 \sqrt{b_{11}}&=&3.37\times10^5 {\rm m/s},\ \  \sqrt{b_{22}}=6.74\times10^5 {\rm m/s},~~~ \\ \sqrt{b_{33}}&=&6.7\times10^4 {\rm m/s},\ \ \ \tilde a_1=0.5,\ \ \ \tilde a_2=0.47. \nonumber
 \end{eqnarray}
Let the magnetic field lie in the $p_1$-$p_2$ plane and the angle $\phi$ determine its direction, i.e., ${\bf n}=(\cos\phi,\sin\phi,0)$. The maximal cross section of this Fermi-surface pocket by the plane perpendicular to ${\bf n}$ is determined by the condition $p_z=({\bf s}{\bf n})$ where ${\bf s}$ is given by Eq.~(\ref{6}). It follows from Eqs.~(\ref{11}), (\ref{12}) that $\delta\gamma_L=0$ and the Berry phase is equal to $\pi$ only if $p_z=0$, i.e., if the magnetic field is perpendicular to ${\bf s}$. In this situation, the maximal cross section passes through the Weyl point. Otherwise, the Weyl point lies outside this cross section, $\delta\gamma_L \neq 0$, and the Berry phase of the orbit deviates from $\pi$. At an arbitrary $\phi$, formulas (\ref{11}) and (\ref{14}) give the following expression for $\delta\gamma_L$:
 \begin{eqnarray}\label{16}
 \delta\gamma_L=~~~~~~~~~~~~~~~~~~~~~~~~~~~~~ \\-\frac{\tilde a_1C_1+\tilde a_2C_2}{2\sqrt{\!(\tilde a_1 \tilde a_2C_1\!+\!(1\!-\!\tilde a_1^2)C_2)^2\!+\!(\tilde a_1 \tilde a_2C_2\!+\!(1\!-\!\tilde a_2^2)C_1)^2}}, \nonumber
 \end{eqnarray}
where $C_1=\cos\phi/\sqrt{b_{11}}$, $C_2=\sin\phi/\sqrt{b_{22}}$.
In Fig.~\ref{fig2}, we show  the dependences of $\delta\gamma_L$ and of $p_z=({\bf s}{\bf n})$ on the angle $\phi$. It is seen that the maximal cross section can be essentially shifted relative to the Weyl point, and $\delta\gamma_L$ noticeably differs from zero in this case. Nevertheless, we emphasize again that the offset of the quantum oscillations will remain one and the same for different $\phi$, and it will be determined by $\gamma=0$.


\begin{thebibliography}{}

\bibitem{armit} N.P. Armitage, E.J. Mele, A. Vishwanath, Rev. Mod. Phys. {\bf 90}, 015001 (2018).

\bibitem{LP} E. M. Lifshitz and L. P. Pitaevskii,
{\it Statistical Physics}, part 2: {\it Theory of the
Condensed State}, Pergamon Press, Oxford (1986).

\bibitem{shen} G.P. Mikitik, Yu.V. Sharlai, Fiz. Nizk. Temp.
{\bf 33}, 586 (2007) [Low Temp. Phys. {\bf 33}, 439 (2007)].

\bibitem{Sh} D. Shoenberg, {\it Magnetic Oscillations in
Metals}, Cambridge University Press, Cambridge, England (1984).


\bibitem{m-sh19}
G.P. Mikitik, Yu.V. Sharlai, J. Low Temp. Phys.
{\bf 197}, 272 (2019)


\bibitem{zilb}  G. E. Zil'berman, Zh. Eksp. Teor. Fiz. {\bf 32},
296 (1957) [Sov. Phys. JETP {\bf 5}, 208 (1957)].

\bibitem{azb}  M. Ya. Azbel', Zh. Eksp. Teor. Fiz. {\bf 39},
1276 (1960) [Sov. Phys. JETP {\bf 12}, 891 (1961)].

\bibitem{falk}  L. A. Fal'kovski\u{i}, Zh. Eksp. Teor. Fiz. {\bf 49}, 609 (1965) [Sov. Phys. JETP {\bf 22}, 423 (1966)].

\bibitem{slut}  A. A. Slutskin, Zh. Eksp. Teor. Fiz. {\bf 53},
767 (1967) [Sov. Phys. JETP {\bf 26}, 474 (1968)].

\bibitem{jetp} G.P. Mikitik, Yu.V. Sharlai, Zh. Eksp. Teor. Fiz.
{\bf 114}, 1375 (1998) [JETP {\bf 87}, 747 (1998)].

\bibitem{prl} G.P. Mikitik, Yu.V. Sharlai, Phys.\ Rev.\ Lett.\
{\bf 82}, 2147 (1999).

\bibitem{g1}
G.P. Mikitik, Yu.V. Sharlai, Phys.\ Rev.\ B
{\bf 65}, 184426 (2002);

\bibitem{g2}
G.P. Mikitik, Yu.V. Sharlai, Phys.\ Rev.\ B
{\bf 67}, 115114 (2003);

\bibitem{alex} A. Alexandradinata, C. Wang, W. Duan. L. Glazman, Phys. Rev. X {\bf 8}, 011027 (2018).

\bibitem{fuchs} J.N. Fuchs, F. Piechon, M.O. Goerbig, G. Montambaux, Eur. Phys. J. B {\bf 77}, 351 (2010).

\bibitem{m-sh12}
G.P. Mikitik, Yu.V. Sharlai, Phys.\ Rev.\ B
{\bf 85}, 033301 (2012);

\bibitem{niu} D. Xiao, M.-C. Chang, Q. Niu, Rev. Mod. Phys. {\bf 82}, 1959 (2010).

\bibitem{sol} A. Soluyanov, D. Gresch, Z. Wang, Q. Wu, M. Troyer, X. Dai, and B.A. Bernevig, Nature {\bf 527}, 495 (2015).


\bibitem{m-sh} G.P. Mikitik, Yu.V. Sharlai, Fiz. Nizk. Temp.
{\bf 22}, 762 (1996) [Low Temp. Phys. {\bf 22}, 585 (1996)].

\bibitem{arnold16} F. Arnold, M. Naumann, S.-C. Wu, Y. Sun, M. Schmidt, H. Borrmann, C. Felser, B. Yan, E. Hassinger, Phys.\ Rev.\ Lett. {\bf 117}, 146401 (2016). 

\bibitem{m-sh21} G.P. Mikitik, Yu.V. Sharlai, Fiz. Nizk. Temp.
{\bf 47}, 342 (2021) [Low Temp. Phys. {\bf 47}, 312 (2021)].


\end{thebibliography}
\end{document}